\documentclass{JHEP3}

\usepackage{epsfig}
\usepackage{amsbsy}
\usepackage{varioref}
\usepackage{pifont}

\def\sla#1{\ifmmode%
\setbox0=\hbox{$#1$}%
\setbox1=\hbox to\wd0{\hss$/$\hss}\else%
\setbox0=\hbox{#1}%
\setbox1=\hbox to\wd0{\hss/\hss}\fi%
#1\hskip-\wd0\box1 }

\title{An Effective Higgsless Theory: Satisfying Electroweak Constraints and a Heavy Top Quark}

\author{Roshan Foadi and  Carl Schmidt\\
Department of Physics and Astronomy, Michigan State University\\
East Lansing, MI 48824, USA\\
	E-mail: \email{foadi@pa.msu.edu},
\email{schmidt@pa.msu.edu}}

\abstract{
The main challenge faced by Higgsless models of electroweak symmetry breaking is to reconcile the experimental constraints imposed
by the precision electroweak data and the top quark phenomenology with the unitarity constraints imposed
by longitudinal gauge boson scattering amplitudes.  In this paper we expand on previous work, giving details of how delocalized fermions can be used to adjust the $S$ parameter to zero,
while keeping the $T$ and $U$ parameters naturally suppressed.   We also show that it is possible to obtain the top quark mass, without affecting the delay of unitarity violation of the $W^+ W^-\rightarrow W^+W^-$
scattering amplitude, by separating the mass scales of the fermion sector ($1/R_f$) from that of the gauge sector ($1/R_g$).
The fermion sector scale $1/R_f$ is only weakly constrained by unitarity of the $t\bar{t}\rightarrow W^+ W^-$ scattering amplitude;
thus the ratio $R_g/R_f$ can be quite large, and the top mass can be easily achieved. Anomalous right-handed couplings involving the third generation quarks also avoid
constraints from experimental data if $1/R_f$ is sufficiently large.
}

\preprint{{~MSUHEP--050824}
{~hep-ph/0509071}}

\begin{document}


\section{Introduction}
\label{sec:Intro}

The only piece of the Standard Model (SM) that is still missing is the Higgs boson.  In the SM
its existence is required to break the Electroweak symmetry and to give mass to
the $W$ and $Z$ bosons, as well as to the matter particles.  It also provides interactions that restore unitarity to massive vector boson scattering at high
energies~\cite{Lee:1977eg}.
However, it is quite possible that these effects arise through other dynamics, with different signatures than a neutral scalar boson.  We must be prepared for any alternatives
that nature may have chosen.

Historically, most scenarios of electroweak symmetry breaking that do not involve a Higgs boson
occur through strong dynamics.  The prototypical example of this is
Technicolor~\cite{Weinberg:1975gm,Susskind:1978ms}. More recently,
new avenues of Higgsless electroweak symmetry breaking have been realized, with inspiration
coming from extra-dimensions~\cite{Csaki:2003dt}--\cite{Foadi:2003xa}.
In fact, it has been argued that these new Higgsless theories in
certain limits may be dual to Technicolor~\cite{Arkani-Hamed:2000ds}--\cite{Agashe:2002jx}.
Like Technicolor, the Higgsless models of electroweak
symmetry breaking solve the unitarity problem by the exchange of new massive vector bosons,
although the new states are typically more weakly coupled than the Techni-rhos.  In any event,
Higgsless theories open up new possibilities for physics beyond the standard model, which might
be revealed at the Large Hadron Collider.  More generally, they offer a new framework in which to
investigate those theories where the unitary problem is addressed by vector-boson exchange.

Higgsless theories have been proposed in five warped dimensions~\cite{Csaki:2003dt}--\cite{Cacciapaglia:2005pa},
in five flat dimensions with brane kinetic terms~\cite{Barbieri:2003pr},
in six dimensions~\cite{Gabriel:2004ua}, and in four-dimensional theory space~\cite{Foadi:2003xa}.
Although differing in details, all of these models have certain generic features
and phenomenology~\cite{Birkedal:2004au}.  The
$SU(2)\times U(1)$ electroweak gauge symmetry (with possible multiple copies of the
$SU(2)$'s or $U(1)$'s) is broken explicitly at certain places in the extra (real or theory-space)
dimension, leaving a massless photon and a $W$ and $Z$ as the light vector boson states.
In addition there is a tower of heavy vector bosons.  It is the new contribution from these 
heavy states that delays the onset of unitarity violation. 

The major obstacle that Higgsless theories have faced so far has been in dealing with constraints
from precision electroweak measurements.   Typically, the mixing between the $W$ and $Z$ 
bosons and the tower of heavy vector bosons produces effects which can be interpreted as
a tree-level contribution to the electroweak $S$ parameter.  Whereas the $T$ parameter can
be made naturally small in these models, the $S$ parameter can only be made small
either by making $T$ larger~\cite{Chivukula:2004pk, SekharChivukula:2004mu} or by increasing the scale of the heavy
vector boson tower~\cite{Georgi:2004iy}.  Neither of these solutions is acceptable; in particular,
a mass scale for the heavy vector bosons that is sufficient to satisfy the constraints from the $S$ parameter
is too large to solve the original unitarity problem.

Recently, a new idea has been introduced to address the issues of the $S$ parameter.
By allowing the left-handed fermion fields to have some finite extent in the extra dimensions,
it is possible to cancel the vector boson contributions and to satisfy the electroweak
constraints~\cite{Cacciapaglia:2004rb}--\cite{SekharChivukula:2005xm}.
Although this appears to require fine tuning, it is conceivable that a mechanism may be found 
to make this natural.  In any event, it is useful to see how far this scenario can be carried in
order to produce a consistent alternative theory of the electroweak symmetry breaking sector.

In this paper we extend upon our previous work~\cite{Foadi:2004ps} by looking at the effects of delocalized
fermions in
more detail.  In particular, we concentrate on the problem of incorporating a heavy top quark into
this scenario.  For simplicity we continue to explore these models in the context of the continuum
theory-space model of Ref.~\cite{Foadi:2004ps}, but we emphasize that the qualitative features that we
find should arise in any Higgsless theory with delocalized fermions.

In Section~\ref{sec:model1} we briefly introduce the gauge sector of this theory.  We also
describe, for comparison purposes, a model (I) of the fermion sector with the fermions localized in the
extra dimension.   In Section~\ref{sec:fermions2} we give the details of our new model (II) of the
fermion sector, where the fermions have some leakage into the extra dimension. We analyze the
solutions of the mass equations, and show that  masses of all the light fermions are easily
obtained, except for the top quark.  We also show that by introducing
some simple assumptions in the fermion sector we can suppress flavor-changing neutral
currents, leading to the standard Cabbibo-Kobayashi-Maskawa (CKM) mixing pattern.  

In Section~\ref{sec:constraints} we show that in Model II with delocalized fermions, it is possible to have the
$S,~T~{\rm and}~U$ parameters agree with data.   We then return to the question of
the top quark mass in Section~\ref{sec:top} and show that it can also be obtained by 
introducing an explicit breaking of the five-dimensional Lorentz
symmetry in the fermion kinetic terms.  Since this symmetry is already explicitly broken
by the compactification, as well as in the brane terms, there are no theoretical barriers to
this. 
By allowing an additional scale factor for the fifth-dimensional kinetic term for the fermions, we are
able to separate the scale of the fermion masses from that of the vector boson masses.
This allows us to obtain a realistic mass for the top quark, while simultaneously suppressing
any dangerous right-handed couplings.  However, just as in the vector boson sector, the
scale of heavy fermions cannot be too large without running into unitarity problems.
We address this question in Section~\ref{sec:unitarity}, where we consider the
scattering of polarized top quarks into longitudinal vector bosons ($t_L\bar{t}_L\rightarrow
W^+_LW^-_L$, where $L$ refers to the helicity $\lambda=-$ for the fermions and
$L$ means longitudinal for the $W$'s).  In the standard model this amplitude is unitarized
by Higgs boson exchange.  We show that in our model, it is unitarized by the tower of
$b^\prime$ quarks.  By requiring that the $J=0$ partial wave amplitude remain less than $1/2$
up to some energy scale, we obtain a constraint on the mass scale of the $b^\prime$ states.
We then investigate this constraint in conjunction with the constraint on the
scale of the $W^\prime$ states coming from unitarity of vector boson scattering.

Finally, in Section~\ref{sec:conclusions}, we will offer our conclusions.

\section{Higgsless Theory with Fermions}
\label{sec:model1}

\subsection{Gauge Sector}
\label{subsec:gauge}

We will use the continuum theory-space model of Refs.~\cite{Foadi:2003xa} and \cite{Foadi:2004ps} to
explore the interplay of the fermion sector in a Higgsless theory.
First, we summarize the gauge sector.
It is an $SU(2)$ gauge theory, defined on a fifth-dimensional line segment, 
$0\leq y\leq\pi R$, where the boundary conditions break the gauge symmetry down to $U(1)$
at one end of the interval. The five dimensional action is
\begin{eqnarray}
{\cal S}&=& \int_0^{\pi R}dy\int d^4x
\left[
-{1\over4(\pi R)\hat{g}_5^2}W^{a\,MN}W^a_{MN}
-\delta(y){1\over4g^2}W^{a\,\mu\nu}W^a_{\mu\nu}\right.\nonumber\\
&&\qquad\qquad\qquad\quad\left.-\delta(\pi R-y)
{1\over4g^{\prime2}}W^{3\,\mu\nu}W^3_{\mu\nu}\right]\ ,
\label{eq:5daction}
\end{eqnarray}
where, in this equation, the indices $M,N$ run over the 5 dimensions,
and we impose the Dirichlet Boundary condition, $W^a_\mu=0$, 
at $y=\pi R$ for $a\ne3$.  The boundary kinetic energy term at $y=0$ is defined by
interpreting the $\delta$-function as $\delta(y-\epsilon)$ with 
$\epsilon\rightarrow0^+$ and the fields having Neumann boundary conditions,
$\partial W^a_\mu/\partial y=0$, at $y=0$.  The $\delta$-function
and the field $W^3_\mu$ at $y=\pi R$ should be interpreted similarly.
Note that in the limit of small $g$ and $g^\prime$ the theory looks like an $SU(2)$ gauge theory
and  a $U(1)$ gauge theory, living at the left and right ends of the fifth-dimensional 
interval, respectively.  It is the bulk fields that connect the $SU(2)$ and the $U(1)$ theories,
and transmit the breaking of the theories down to a single $U(1)_{EM}$.

The five-dimensional gauge fields can be expanded in a tower of four-dimensional Kaluza-Klein
(KK) states:
\begin{eqnarray}
W^{\pm\mu}(x,y)&=& \sum_{n=0}^\infty f_n(y)W_n^{\pm\mu}(x)\nonumber\\
W^{3\mu}(x,y)&=& eA^\mu(x)+\sum_{n=0}^\infty g_n(y)Z_n^{\mu}(x)\ ,
\label{eq:gaugeKK}
\end{eqnarray}
where $W_n^{\pm\mu}$ has mass $m_{W_n}$, and
$Z_n^{\mu}$ has mass $m_{Z_n}$. 
The lowest states of the tower, 
$W_0^{\pm\mu}$ and $Z_0^\mu$, will be identified as the standard model $W^\pm$
and $Z$ bosons, respectively.   
The masses satisfy the equations
\begin{eqnarray}
\hat{m}_{W_n}\tan \hat{m}_{W_n} & = & \lambda^2 \nonumber \\
\left(\hat{m}_{Z_n}-{\lambda^2\lambda'^2\over \hat{m}_{Z_n}}\right)\tan \hat{m}_{Z_n}
& = & \lambda^2+\lambda'^2 \ ,
\label{eq:gaugemasseq}
\end{eqnarray}
where $\lambda\equiv g/\hat{g}_5$, and $\lambda'\equiv g'/\hat{g}_5$, and we have
introduced the notation $\hat{m}\equiv m\pi R$ for dimensionless scaled masses.
Solving for the masses perturbatively in
$\lambda^2$ and $\lambda'^2$, we obtain
\begin{eqnarray}
m_W^2\equiv m_{W_0}^2 & = & {\lambda^2\over (\pi R)^2}\left[1-{\lambda^2\over 3}
+{\cal O}(\lambda^4)\right] \nonumber \\
m_Z^2\equiv m_{Z_0}^2 & = & {\lambda^2+\lambda'^2\over (\pi R)^2}\left[1-{\lambda^2+\lambda'^2\over 3}
+{\lambda^2\lambda'^2\over\lambda^2+\lambda'^2}+{\cal O}(\lambda^4)\right], \
\label{eq:gaugemassSM}
\end{eqnarray}
for the standard model gauge bosons, and
\begin{eqnarray}
m_{W_n}^2 & = & \left({n\over R}\right)^2\left[1+2{\lambda^2\over (\pi n)^2}
+{\cal O}(\lambda^4)\right] \nonumber \\
m_{Z_n}^2 & = & \left({n\over R}\right)^2\left[1+2{\lambda^2+\lambda'^2\over (\pi n)^2}
+{\cal O}(\lambda^4)\right], \
\label{eq:gaugemassHeavy}
\end{eqnarray}
for the heavy gauge bosons.

Solving the equations of motion and applying the 
boundary conditions in
the fifth dimension, gives the following gauge boson wave functions:
\begin{eqnarray}
f_n(y) & = & F_n \sin\left[m_{W_n}(\pi R-y)\right] \nonumber \\
g_n(y) & = & G_n \left[\cos(m_{Z_n}y)-\frac{\hat{m}_{Z_n}}{\lambda^2}\sin(m_{Z_n}y)\right] \,
\label{eq:gaugefunctions}
\end{eqnarray}
Requiring the KK states to be canonically normalized in the four-dimensional
Lagrangian, and using Eqs.~(\ref{eq:gaugemasseq}) to simplify,
leads to the following expressions for the normalization constants:
\begin{eqnarray}
F_n & = & \hat{g}_5\left[{1\over2}\left(1+{\sin2\hat{m}_{W_n}\over2\hat{m}_{W_n}}\right)\right]^{-1/2}
\nonumber \\
G_n & = & \hat{g}_5\left[{1\over2}\left(1-{\sin2\hat{m}_{Z_n}\over2\hat{m}_{Z_n}}\right)
+{1\over\lambda^2}\sin^2\hat{m}_{Z_n}+{\hat{m}_{Z_n}^2\over2\lambda^4}
\left(1+{\sin2\hat{m}_{Z_n}\over2\hat{m}_{Z_n}}\right)\right]^{-1/2}\ .
\label{eq:gaugenorm}
\end{eqnarray}
Expanding perturbatively in $\lambda^2$ and $\lambda'^2$ gives
\begin{eqnarray}
F_0 & = & {g\over\lambda}\left[1+{\lambda^2\over 6}+{\cal O}(\lambda^4)\right] \nonumber \\
F_n & = & {\sqrt{2} g\over\lambda}\left[1-{\lambda^2\over 2(n\pi)^2}+{\cal O}(\lambda^4)\right]
\ n=1,2,....\nonumber \\
G_0 & = & {g\lambda\over\sqrt{\lambda^2+\lambda'^2}}\left[1-{\lambda^4+2\lambda^2\lambda'^2-2\lambda'^4\over
6\left(\lambda^2+\lambda'^2\right)}+{\cal O}(\lambda^4)\right]\nonumber \\
G_n & = & {\sqrt{2}g\lambda\over n\pi}\left[1-{3\over 2}{\lambda^2+\lambda'^2\over (n\pi)^2}
+{\cal O}(\lambda^4)\right]\ n=1,2,....\ .
\label{eq:gaugenormpert}
\end{eqnarray}

\subsection{Fermion Sector: Model I}
\label{subsec:fermion1}

It is useful to consider first a model with localized fermions, for comparison purposes.
The continuum limit of the fermion action considered in ref.~\cite{Foadi:2003xa} can be written
\begin{eqnarray}
{\cal S}^{(I)}&=& \int_0^{\pi R}dy\int d^4x
\left[
\delta(y)i\bar{\psi}_L\sla{D}\psi_L+\delta(\pi R-y)i\bar{\psi}_R\sla{D}\psi_R
\right]\ ,
\label{eq:fermionaction1}
\end{eqnarray}
where the covariant derivatives on the left- and right-handed fields are
\begin{eqnarray}
\sla{D}\psi_L&=& \left(\sla{\partial}-iT^a\sla{W}^a(y)-iY_L\sla{W}^3(\pi R)\right)\psi_L\nonumber\\
\sla{D}\psi_R&=& \left(\sla{\partial}-iY_R\sla{W}^3(y)\right)\psi_R\ .
\label{eq:derivs1}
\end{eqnarray}
Note that the left-handed field $\psi_L$ lives at $y=0$ 
but couples also to the gauge field $W^3$ at $y=\pi R$.  This non-locality is perfectly 
natural from the standpoint of a continuum theory-space model.   In the theory-space
interpretation $W^a(0)$ and $W^3(\pi R)$ are just the gauge fields for independent $SU(2)$
and $U(1)$ gauge groups, and $y$ is a (continuous) label for the independent gauge groups.

To give the fermions a mass in this model, a nonlocal mass term involving a Wilson line 
running between the two branes must be included.  However, the most critical problem
with this fermion action is that it produces electroweak radiative 
corrections that are too large.  Therefore, we are led to a theory with delocalized fermions.

\section{Fermion Sector: Model II}
\label{sec:fermions2}

Drawing on the analogy of the gauge action (\ref{eq:5daction}), which has $SU(2)$ and $U(1)$
kinetic terms peaked at the two ends of the interval and connected through the bulk kinetic
term, we considered in Ref.~\cite{Foadi:2004ps}  a theory with left-handed and right-handed fermion
kinetic terms peaked at the two ends of the
interval and connected through a bulk fermion kinetic term.  The fermion action is
\begin{eqnarray}
{\cal S}^{(II)} = \int_0^{\pi R}dy && \int d^4x
\left[{1\over\pi R}\left({i\over2}\bar{\psi}\Gamma^M D_M\psi + \ {\rm h.c.}
-M\bar{\psi}\psi\right)\right.\nonumber\\
&&\left.
+\delta(y){1\over t_L^2}i\bar{\psi}_L\sla{D}\psi_L
+\delta(\pi R-y)\left({1\over t_{u_R}^2}i\bar{u}_R\sla{D}u_R
+{1\over t_{d_R}^2}i\bar{d}_R\sla{D}d_R\right)
\right]\ ,
\label{eq:fermionaction2}
\end{eqnarray}
The five-dimensional Dirac matrices are defined in terms of the four-dimensional ones by
$\Gamma^M = (\gamma^\mu,-i\gamma^5)$.  The five-dimensional fermion is equivalent to
a four-dimensional Dirac fermion, $\psi=\psi_L+\psi_R$, where $\psi_L$ and $\psi_R$ are $SU(2)$
doublets, $\psi_L=(u_L,d_L)$ and $\psi_R=(u_R,d_R)$. 
We have written the action for one doublet, consisting of an up and a down quark.  We will
discuss the possibility of more generations and mixing in Section~\ref{subsec:mixing}.
We can assume the bulk mass $M$ to be real in Eq.~(\ref{eq:fermionaction2}) without any loss of generality.
Any imaginary part of $M$
can be removed by the replacement $\psi\rightarrow e^{(i{\rm Im} My)}\psi$.  
The sign of $M$, however, is physical.
The boundary kinetic energy term at $y=0$ is defined by
interpreting the $\delta$-function as $\delta(y-\epsilon)$ for $\epsilon\rightarrow0^+$
with the boundary condition
$\psi_R=0$ at $y=0$.  Similarly, the boundary  term at $y=\pi R$ is defined by
interpreting the $\delta$-function as $\delta(\pi R-y+\epsilon)$ with the boundary condition
$\psi_L=0$ at $y=\pi R$.  The general treatment of possible fermion boundary conditions
can be found in Ref.~\cite{Csaki:2003sh}.

The covariant derivative in Eq.~(\ref{eq:fermionaction2}) is
\begin{equation}
D_M\psi = \left(\partial_M - iT^aW_M^a(y)-iY_LW_{M}^3(\pi R)\right)\psi,
\label{eq:bulkderiv}
\end{equation}
where $Y_L$ is the $\psi_L$ hypercharge. At the interval
ends the four-dimensional part of the covariant derivative (\ref{eq:bulkderiv}) becomes:
\begin{eqnarray}
(\sla{D}\psi_L)_{y=0} & = & \left(\sla{\partial}-iT^a\sla{W}^a(0)-iY_L\sla{W}^3(\pi R)\right)\psi_L \nonumber \\
(\sla{D}\psi_R)_{y=\pi R} & = & \left(\sla{\partial}-iT^3\sla{W}^3(\pi R)-iY_L\sla{W}^3(\pi R)\right)\psi_R \nonumber \\
& = & \left(\sla{\partial}-iY_R\sla{W}^3(\pi R)\right)\psi_R \ ,
\label{eq:derivends}
\end{eqnarray}
where the $\psi_R$ hypercharge, $Y_R$, is related to $Y_L$ by $Y_R=T^3+Y_L$, as in the SM. Note that 
$Y_R$ is a 2$\times$2 diagonal matrix, with the $u_R$ hypercharge on the upper left, 
and the $d_R$ hypercharge on the
lower right. Therefore, at $y=\pi R$ the covariant derivative term, $\bar{\psi}_R\sla{D}\psi_R$, 
splits into two 
separately gauge invariant terms, $\bar{u}_R\sla{D}u_R$ and $\bar{d}_R\sla{D}d_R$, as in 
Eq.~(\ref{eq:fermionaction2}).  Note also that in the limit of small $t_L$, $t_{u_R}$, and $t_{d_R}$ 
 the action ${\cal S}^{(II)}$ describes massless left-handed fermions gauged under an
 $SU(2)\times U(1)$ group living on the left end of the fifth-dimensional interval, and 
massless right-handed fermions gauged under a $U(1)$ living on the right end of the interval, 
exactly as in model I.  
It is the presence of the bulk fields which allow these light states to communicate with each
other, supplying the analog of the Yukawa coupling of the SM, and giving mass to the fermions.

\subsection{Fermion Masses and Wave Functions}
\label{subsec:fermionmass}
Let $\chi$ denote either $u$, the up-type fermions, or $d$, the down-type fermions.
The fermion fields can be expanded in a tower of four-dimensional KK states:
\begin{eqnarray}
\chi_L(x,y)&=& \sum_{n=0}^\infty \alpha_n(y)\chi_{n_L}(x)\nonumber\\
\chi_R(x,y)&=& \sum_{n=0}^\infty \beta_n(y)\chi_{n_R}(x) \ .
\label{eq:fermionKK}
\end{eqnarray}
The four dimensional fields $\chi_{n_L}$ and $\chi_{n_R}$ are
the left-handed and right-handed projection, respectively, of a mass-$m_n$ Dirac fermion,
$\chi_n=\chi_{n_L}+\chi_{n_R}$. Wave-function and mass equations are obtained by diagonalizing the free-field action.
Turning off the gauge couplings, the action of Eq.~(\ref{eq:fermionaction2}) becomes
${\cal S}_u^{\rm free} + {\cal S}_d^{\rm free}$, where
\begin{eqnarray}
{\cal S}_\chi^{\rm free} = \int_0^{\pi R} dy\int d^4x
&& \left[{1\over\pi R}\left(i\bar{\chi}_L\sla{\partial}\chi_L 
+ i\bar{\chi}_R\sla{\partial}\chi_R\right.\right. \nonumber \\
&&\left.
- {1\over 2}\left(\bar{\chi}_R\partial_5 \chi_L-\bar{\chi}_L\partial_5 \chi_R + \ {\rm h.c.}\right)
-M\left(\bar{\chi}_R \chi_L
+\bar{\chi}_L \chi_R\right)\right) \nonumber \\
&&\left.
+\delta(y-\epsilon) {1\over t_L^2}i\bar{\chi}_L\sla{\partial}\chi_L
+\delta(\pi R-\epsilon-y) {1\over t_{\chi_R}^2}i\bar{\chi}_R\sla{\partial}\chi_R\right]\ .
\label{eq:freeaction}
\end{eqnarray} 

In Eq.~(\ref{eq:freeaction}) we have explicitly included a finite $\epsilon$ to
push the delta-function terms slightly away from the interval ends, allowing us to
unambiguously impose the boundary conditions
\begin{eqnarray}
\chi_R & = & 0 \ {\rm at} \ y=0 \nonumber \\
\chi_L & = & 0 \ {\rm at} \ y=\pi R \ .
\label{eq:BC}
\end{eqnarray}
The field equations in the bulk can be obtained by variation of ${\cal S}_\chi^{\rm free}$.
Integrating these equations around the delta-functions, 
taking the limit $\epsilon\rightarrow 0$,  
and using the boundary conditions (\ref{eq:BC}), leads to alternative expressions for the boundary
limits:\footnote{Comparing Eq.~(\ref{eq:BC}) with
Eq.~(\ref{eq:limits}), we see that $\chi_R$ has 
a discontinuity at $y=0$, but $\chi_L$ is continuous.  Similarly, $\chi_L$ is discontinuous
at $y=\pi R$, but $\chi_R$ is continuous.}
\begin{eqnarray}
\lim_{y \rightarrow 0^+} \chi_R & = & -{1\over t_L^2}i\sla{\partial}\chi_L(0) \nonumber \\
\lim_{y \rightarrow \pi R^-} \chi_L & = & -{1\over t_{\chi_R}^2}i\sla{\partial}\chi_R(\pi R) \ .
\label{eq:limits}
\end{eqnarray}
It is now most convenient to treat the action as finite in the bulk ($0<y<\pi R$) with
Eq.~(\ref{eq:limits}) as boundary conditions. In this case the bulk equations of motion become
\begin{eqnarray}
\dot{\alpha_n} + M\alpha_n - m_n\beta_n & = & 0 \nonumber \\
\dot{\beta_n} - M\beta_n + m_n\alpha_n  & = & 0 \ ,
\label{eq:wave}
\end{eqnarray}
with the boundary conditions
\begin{eqnarray}
\beta_n(0) &=& -{\hat{m}_n\over t_L^2}\alpha_n(0) \nonumber \\
\alpha_n(\pi R) &=& -{\hat{m}_n\over t_{\chi_R}^2}\beta_n(\pi R)\ .
\label{eq:BCwave}
\end{eqnarray}
(Recall that masses with a hat are expressed in units of  $(\pi R)^{-1}$; {\it i.e.,}
$\hat{m}\equiv m\pi R$.)
The solutions to these equations are simplest in the case of zero bulk mass $M$.  We
study this case first and then look at the numerical solutions for nonzero $M$.

\paragraph{(i) $M=0$.}
With no bulk mass, the solutions of Eqs.~(\ref{eq:wave}) are
\begin{eqnarray}
\alpha_n & = & A_n\left[\cos(m_n y) - {\hat{m}_n\over t_L^2}\sin(m_n y)\right] \nonumber \\
\beta_n & = & -A_n\left[{\hat{m}_n\over t_L^2}\cos(m_n y)+\sin(m_n y)\right] \ .
\label{eq:wavefunctions}
\end{eqnarray}
Applying the boundary conditions given in Eqs.~(\ref{eq:BCwave}) to these solutions leads to an equation
for the fermion masses,
\begin{equation}
(t_L^2 + t_{\chi_R}^2)\hat{m}_n\tan \hat{m}_n
+ \hat{m}_n^2 = t_L^2t_{\chi_R}^2 \ .
\label{eq:massequation}
\end{equation}
The lowest mass state of the KK tower corresponds to a standard model fermion.  This
light mass can be easily obtained in a perturbation expansion if we assume $t_L^2$ to
be small:
\begin{equation}
\hat{m}_0 = {t_L t_{\chi_R}\over\sqrt{1+t_{\chi_R}^2}}\left[1-{1+t_{\chi_R}^2
+t_{\chi_R}^4/3\over 2\left(1+t_{\chi_R}^2\right)^2}t_L^2 + {\cal O}(t_L^4)\right] \ .
\label{eq:mass}
\end{equation}
If we also assume $t_{\chi_R}^2$ to be small, the heavy state masses are
\begin{equation}
\hat{m}_n=\pi\left(n-{1\over 2}\right)\left[1+{t_L^2+t_{\chi_R}^2\over\pi^2\left(n-{1\over 2}\right)^2}
+{\cal O}(t^4)\right];\ \ n=1,2,...\ .
\label{eq:mass_n}
\end{equation}

The normalization factor $A_n$ can be fixed by requiring that the KK states be canonically normalized in
the four-dimensional Lagrangian.  We obtain
\begin{equation}
A_n \  = \ \left[{1\over2}\left(1+{\sin2\hat{m}_{n}\over2\hat{m}_{n}}\right)
+{1\over t_L^2}\cos^2\hat{m}_{n}+{\hat{m}_{n}^2\over2t_L^4}
\left(1-{\sin2\hat{m}_{n}\over2\hat{m}_{n}}\right)\right]^{-1/2}\ .
\label{eq:gaugenormFermions}
\end{equation}
For $t_L^2$ small this gives
\begin{equation}
A_0=t_L\left[1-{1+t_{\chi_R}^2
+t_{\chi_R}^4/3\over 2\left(1+t_{\chi_R}^2\right)^2}t_L^2 + {\cal O}(t_L^4)\right],
\label{eq:coefficients}
\end{equation}
for the lightest state, and for both $t_L^2$ and $t_{\chi_R}^2$ small this gives
\begin{equation}
A_n={\sqrt{2}t_L^2\over\pi\left(n-{1\over 2}\right)}\left[1-
{3\over 2}{t_L^2+t_{\chi_R}^2\over\pi^2\left(n-{1\over 2}\right)^2}+{\cal O}(t^4)\right];\ \ n=1,2,...,
\label{eq:coefficients_n}
\end{equation}
for the heavy states.

From Eq.~(\ref{eq:mass}) we see that the lightest fermion mass is suppressed by the factor 
$t_L t_{\chi_R}$. For small values of these parameters this lightest Dirac fermion lies mainly
on the branes, with small contribution from the bulk.  The left-handed bulk wave function $\alpha_0(y)$ goes
to zero as $t_L\rightarrow 0$, and the right-handed bulk wave function $\beta_0(y)$ goes to zero as
$t_R\rightarrow 0$. Since the fermion masses arise from the $\partial_5$-terms, which mix left-handed
and right-handed wave functions, it follows that $m_0$ goes to zero, as either $t_L\rightarrow 0$ or
$t_R\rightarrow 0$. 

Note also that
Eq.~(\ref{eq:mass}) and Eq.~(\ref{eq:mass_n}) are symmetric in $t_L$ and $t_{\chi_R}$. This was expected, since
the mass equation is
$t_L-t_{\chi_R}$ symmetric. However, we shall treat $t_L$ and $t_{\chi_R}$ differently.
For starters, $t_L$ is an SU(2) invariant parameter, whereas $t_{\chi_R}$ can have
different values for the up and down fermions.  We shall take this distinction further by assuming
that $t_L$ is universal for all quarks and leptons, and that the different particle masses are
determined by $t_{\chi_R}$.  We shall find in section \ref{sec:constraints} that if $t_L$ is of order
$\lambda$, it can be used to cancel the positive contribution to the $S$-parameter
that comes from the gauge sector.    To have an idea of the orders of magnitude
involved, let us assume $R^{-1}\sim\ 1$ TeV and $t_L\sim\lambda\sim\ 10^{-1}$.  
Then $t_{\chi_R}$ ranges from $\sim 10^{-11}$ for the lightest neutrino,
to $\sim 10^{-2}$ for the charm quark.  We shall return to the issue of the third generation
quarks shortly.

\begin{figure}
\includegraphics[width=6in,height=4.8in]{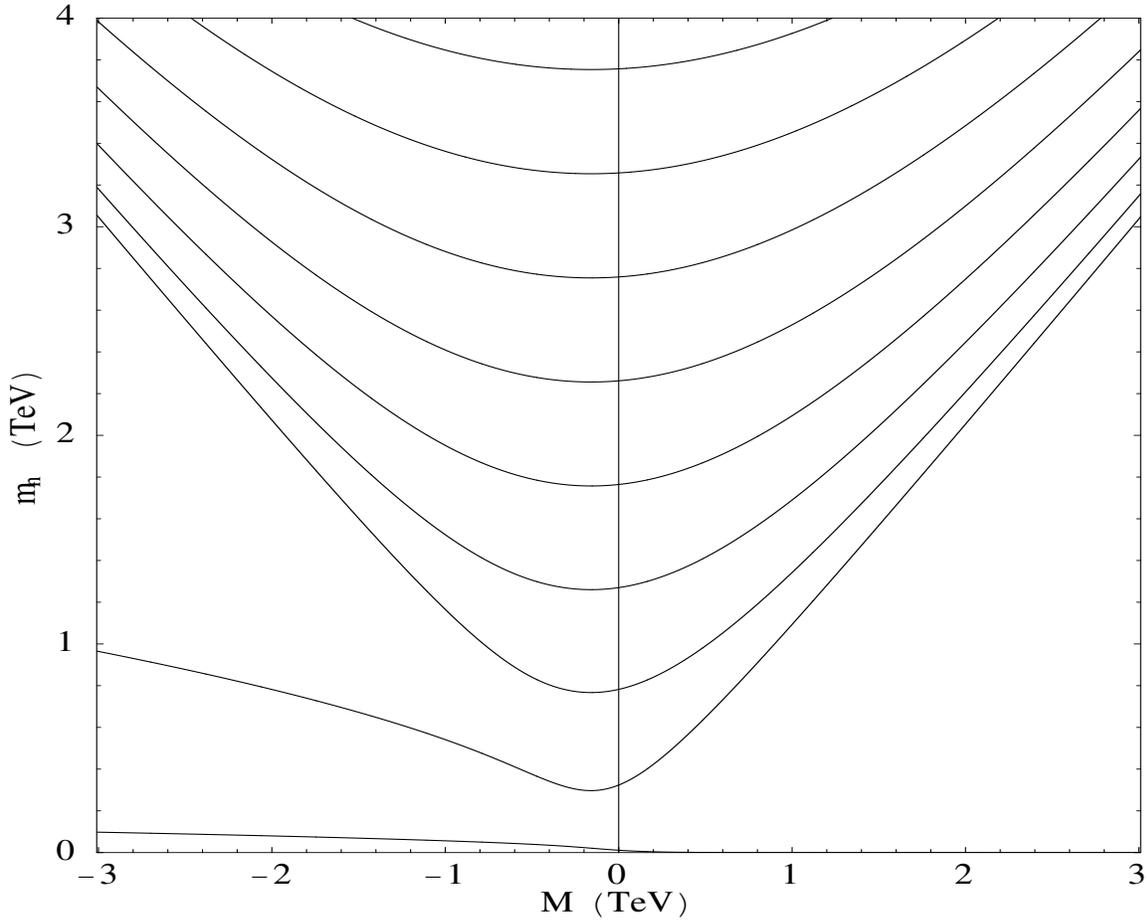}
\caption{Masses of fermions, as a function of the bulk mass,
for $t_L=10^{-1}$, $t_R=1$, and $1/R=500$ GeV. }
\label{fig:mass}
\end{figure}

\paragraph{(ii) $M\neq 0$.}
For nonzero bulk mass, the analysis is similar; the equations are just a bit longer.
The equation for the fermion masses becomes
\begin{equation}
\left[ \left(\hat{m}_n^2+t_L^2 t_{\chi_R}^2\right)\hat{M}+\left(t_L^2+t_{\chi_R}^2\right)\hat{m}_n^2\right]
T(m_n)\ =\ t_L^2 t_{\chi_R}^2-\hat{m}_n^2\ ,
\label{eq:massequation2}
\end{equation}
where the function $T({m}_n)$ depends on the relation between $m_n$ and $M$:
\begin{equation}
T(m_n)\ =\ \left\{\begin{array}{lll}
{\tan{\sqrt{\hat{m}_n^2-\hat{M}^2}}\over\sqrt{\hat{m}_n^2-\hat{M}^2}}&\qquad\mbox{for}
&\qquad m_n>|M|\\
&&\\
{\tanh{\sqrt{\hat{M}^2-\hat{m}_n^2}}\over\sqrt{\hat{M}^2-\hat{m}_n^2}}&\qquad\mbox{for}
&\qquad |M|>m_n\ .\\
\end{array}
\right.
\end{equation}
For a large and positive bulk mass $\hat{M}>0$, $\hat{M}\gg1$, there is one light solution to this
mass equation, given approximately by
\begin{equation}
\hat{m}_0^2\ \approx\ t_L^2t_{\chi_R}^2e^{-2\hat{M}}\ .
\label{eq:mass2}
\end{equation}
For a large and negative bulk mass $\hat{M}<0$, $|\hat{M}|\gg1$, there are two light solutions,
which are asymptotically given by (for $t_L<t_{\chi_R}$)
\begin{eqnarray}
\hat{m}_{0}^2&\approx&2|\hat{M}|t_L^2\nonumber\\
\hat{m}_{1}^2&\approx&2|\hat{M}|t_{\chi_R}^2\ .
\label{eq:mass2p}
\end{eqnarray}
This behavior is displayed in Fig.~\ref{fig:mass}, where we plot the mass
states as a function of the bulk mass $M$, with the other parameters fixed at $t_L=10^{-1}$,
$t_{\chi_R}=1$, and $1/R=500$ GeV.  The transformation $M\rightarrow -M$ can be shown to
be equivalent to a reflection in the fifth dimension.  Since the boundary conditions (\ref{eq:BC})
that we have imposed are asymmetric in this reflection, we obtain the asymmetric behavior
in $M\rightarrow -M$ of Fig.~\ref{fig:mass}, even in the absence of the brane kinetic terms.  
The heavier modes are less asymmetric, because they are less affected by the boundaries.

\subsection{Multiple Generations and Fermion Mixing}
\label{subsec:mixing}
It is not difficult to implement multiple generations in our fermion model II .  In general the bulk
mass, $M$, and the normalizations of the brane kinetic terms, $t_L^{-2}$,
$t_{u_R}^{-2}$, and $t_{d_R}^{-2}$, would be independent $3\times3$ matrices for both
the leptons and the quarks.  However, this proliferation of mixing matrices would open
the door to large flavor-changing neutral currents, which must somehow be avoided.  
The simplest way to achieve this is to restrict all of the flavor physics to the right brane,
and impose a global $U(3)_{\rm quark}\times U(3)_{\rm lepton}$ symmetry on the quark 
and lepton doublets in the bulk and on the left brane.  This flavor symmetry would only be 
broken by the kinetic terms on the right brane (which, incidentally, is also the only place where the
$SU(2)$ weak gauge symmetry is broken).   The generalization of the fermion action
in Eq.~(\ref{eq:fermionaction2}) is
\begin{eqnarray}
{\cal S}^{(II)} &=& \int_0^{\pi R}dy \int d^4x
\left[{1\over\pi R}\left({i\over2}\bar{\psi}^i\Gamma^M D_M\psi^i + \ {\rm h.c.}
-M\bar{\psi}^i\psi^i\right)\right.\nonumber\\
&&\left.\qquad
+\delta(y){1\over t_L^2}i\bar{\psi}_L^i\sla{D}\psi_L^i
+\delta(\pi R-y)\left(i\bar{u}_R^iK_u^{ij}\sla{D}u_R^j
+i\bar{d}_R^{i\prime}K_d^{ij}\sla{D}d_R^{j\prime}\right)\right] \ ,
\label{eq:fullaction}
\end{eqnarray}
where $i$ and $j$ are generation indices, and there is an equivalent contribution
for leptons.   In principle the $t_L$ and $M$ parameters, as well as the $K$ matrices,
 can be different for the lepton and quark sectors.
The five-dimensional fermion fields $\psi^i$'s can be considered four-dimensional Dirac fermions, which
are also $SU(2)$ doublets:
\begin{eqnarray}
\psi^i=\psi_L^i+\psi_R^i=\left( \begin{array}{c}
u_L^i \\
d_L^{i'}
\end{array} \right)+\left( \begin{array}{c}
u_R^i \\
d_R^{i'}
\end{array} \right)
\label{eq:doublets}
\end{eqnarray}

The quark sector matrices $K_u$ and $K_d$ are arbitrary Hermitian matrices; however,
we can exploit the $U(3)_{\rm quark}$ symmetry of the quark fields in the bulk and on the
left brane to reduce the number of real physical parameters to $9\, +\, 9\, -\, (9-1)=10$, where we have
taken into account the fact that the $U(1)$ part of $U(3)$ is just an overall phase symmetry.
We can identify these 10 parameters as the six quark masses, and the four physical 
parameters of the
CKM matrix.
To see how this works, we first perform an $SU(3)$ transformation on the $\psi^i$ to diagonalize $K_u$.
Thus, without loss of generality, we can assume $K^{ij}_u=(t_{u_{iR}}^{-2})\delta^{ij}$.
We can also assume that $K^{ij}_d$ is diagonalized by a unitary matrix $V$, so that
$K^{ij}_d=V^{ik}(t_{d_{kR}}^{-2})(V^\dagger)^{kj}$.  We now relate the (primed) 
gauge eigenstates to the (unprimed) mass eigenstates by the redefinition
$d_{L,R}^{i\prime}\ =\ V^{ij}d_{L,R}^j$.
The action now becomes
\begin{eqnarray}
{\cal S}^{(II)} &=& \int_0^{\pi R}dy  \int d^4x
\left[{1\over\pi R}\left({i\over2}\bar{\psi}^i\Gamma^M D_M\psi^i + \ {\rm h.c.}
-M\bar{\psi}^i\psi^i\right)\right.\nonumber\\
&&\left.\qquad
+\delta(y){1\over t_L^2}i\bar{\psi}_L^i\sla{D}\psi_L^i
+\delta(\pi R-y)\left({1\over t_{u_{iR}}^2}i\bar{u}_R^i\sla{D}u_R^i
+{1\over t_{d_{iR}}^2}i\bar{d}_R^{i}\sla{D}d_R^{i}\right)\right] \ ,
\label{eq:fullaction2}
\end{eqnarray}
where
\begin{eqnarray}
\psi_L^i=\left( \begin{array}{c}
u_L^i \\
V^{ij}d_L^j
\end{array} \right) \,,\qquad
\psi_R^i=\left( \begin{array}{c}
u_R^i \\
V^{ij}d_R^j
\end{array} \right) \ .
\label{eq:doublets2}
\end{eqnarray}
The unitary matrix $V$ corresponds precisely to the CKM matrix in the SM, only
arising in terms that involve the exchange of charged $SU(2)$ gauge bosons.  Just as for the
CKM matrix, it can be reduced to three real parameters and one phase, via five
independent phase redefinitions of the $u_{L,R}$ and $d_{L,R}$ fields.

It is not difficult to see that any implementations of the SM
can be mapped into this picture.
In the lepton sector, for example, we could induce a see-saw mechanism by including a Majorana mass term for
the neutrino, at $y=\pi R$: in that case the matrix $V$ would contain two more physical parameters, corresponding
to the Majorana phases of the MNS matrix.  Alternatively, we could have a zero-mass neutrino, by imposing the
boundary condition $\nu_R=0$ at $y=\pi R$. In that case the number of physical parameters would be 9-(9-3)=3,
corresponding to the
three lepton masses.

\section{Electroweak Constraints}
\label{sec:constraints}

In Ref.~\cite{Foadi:2003xa} we showed that the electroweak constraints in model I could be parametrized to order $\lambda^2$ in terms of the oblique parameters $S$, $T$, and $U$~\cite{Peskin:1991sw}.  In Ref.~\cite{Foadi:2004ps} we found that this result also applies to model II
if we assume that $t_L\sim\lambda$ and we neglect the contribution of $t_{\chi_R}$.  In this section we shall review these results, filling in some of the details for both models.
We also extend these results by 
 investigating the effects of a small but nonzero $t_{\chi_R}$.  

We begin by discussing the couplings of the light fermions to the tower of vector bosons. 
The simplest coupling to obtain is just that of the massless photon.  Normalizing the four-dimensional photon field, we obtain a relation for the electromagnetic coupling that is
valid in all of the fermion models considered in this paper:
\begin{eqnarray}
e^2&=& \left({1\over \hat{g}_5^2}+{1\over g^2}+{1\over g^{\prime2}}\right)^{-1}\ .
\label{eq:em1}
\end{eqnarray}
This relation holds, independent of how the fermions are incorporated in the model, due to the fact that the photon wave function is flat.

The couplings of the light fermions to the charged and neutral vector bosons can be
expressed in terms of the following interaction Lagrangians
\begin{eqnarray}
{\cal L}_{CC}&=& {W^{+\mu}_n\over\sqrt{2}}\bar{u}\left[g^{CC}_{Ln}
\gamma_\mu P_L+g^{CC}_{Rn}
\gamma_\mu P_R\right]d\ +\ {\rm h.c.}\nonumber\\
{\cal L}_{NC}&=& Z^{\mu}_n\,\bar{\psi}
\left[g^{NC}_{Ln}T^3\gamma_\mu P_L+g^{NC}_{Rn}T^3\gamma_\mu P_R
+g^{NC}_{Qn}Q\gamma_\mu 
\right]\psi\ +\ {\rm h.c.}\ .
\label{eq:CCLag}
\end{eqnarray}
Let us investigate these couplings for both models.

\paragraph{(i) Couplings in Model I.}

In model I with brane-localized fermions the couplings are determined by the values of the light boson wavefunctions at the boundaries.  The right-handed couplings are both zero,
\begin{eqnarray}
g^{CC(I)}_{Rn}\ =\ g^{NC(I)}_{Rn}\ =\ 0\ ,
\label{eq:couplingsWF1r}
\end{eqnarray}
while the remaining couplings are given by
\begin{eqnarray}
g^{CC(I)}_{Ln}&=& f_n(0)\nonumber\\
g^{NC(I)}_{Ln}&=&g_n(0)-g_n(\pi R)\nonumber\\
g^{NC(I)}_{Qn}&=& g_n(\pi R)\ .
\label{eq:couplingsWF1}
\end{eqnarray}

It is not difficult to show that the couplings of the fermions to all  of the heavier vector boson
 states in model I are suppressed by a factor of 
$m_W/m_{W_n}\approx \lambda/(n\pi)$, relative
to their couplings to the standard model $W$ and $Z$.  We find for $n\ge1$
\begin{eqnarray}
g^{CC(I)}_{Ln}&=& g\left({\sqrt{2}m_W\over m_{W_n}}\right)
\left[1+{\cal O}(\lambda^2)\right]\nonumber\\
g^{NC(I)}_{Ln}&=& g\left({\sqrt{2}m_W\over m_{W_n}}\right)
{g^2+(-1)^ng^{\prime2}\over g^2}
\left[1+{\cal O}(\lambda^2)\right]\nonumber\\
g^{NC(I)}_{Qn}&=& -g\left({\sqrt{2}m_W\over m_{W_n}}\right)(-1)^n
{g^{\prime2}\over g^2}
\left[1+{\cal O}(\lambda^2)\right]\ .
\label{eq:CCKK1}
\end{eqnarray}
This implies that the contribution of the heavier vector bosons to
four-fermion operators at zero-momentum transfer are suppressed by a factor of $\lambda^4$
relative to the standard model $W$ and $Z$ contributions.  (There is a relative factor of
$\lambda^2$ from the couplings and an additional factor of $\lambda^2$ due to 
the heavy vector boson masses in the boson propagator.)  Therefore, to order $\lambda^2$,
we only need to consider the couplings to the standard model $W$ and $Z$ for electroweak
precision measurements.

We have now seen in model I that only the couplings $g^{CC}_{L0}$, $g^{NC}_{L0}$,
and $g^{NC}_{Q0}$ are relevant to electroweak precision measurements at order $\lambda^2$. Therefore,  the deviations from the standard model can be parametrized in terms the oblique parameters $S$, $T$, and $U$~\cite{Peskin:1991sw}.  Any additional universal parameters, such as those considered in Ref.~\cite{Barbieri:2004qk} can be neglected at this order.   Following Ref.~\cite{Burgess:1993vc}, the $S$, $T$, and $U$ parameters appear as
deviations in relations of the universal couplings of the fermions to the $W$ and $Z$
gauge bosons.  Taking $\alpha$, $m_W$, and $m_Z$ as the
fundamental input observables, we find
\begin{eqnarray}
g^{CC}_{L0}&=& {e\over s}\left[1+{\alpha S\over4s^2}-
{c^2\alpha T\over2s^2}-{(c^2-s^2)\alpha U\over8s^4}\right]\nonumber\\
g^{NC}_{L0}&=& {e\over sc}\left[1+{\alpha S\over4s^2}-
{(c^2-s^2)\alpha T\over2s^2}-{(c^2-s^2)\alpha U\over8s^4}\right]
\nonumber\\
g^{NC}_{Q0}&=& -{es\over c}\left[1+{\alpha T\over2s^2}+{\alpha U\over8s^4}\right]
\ ,
\label{eq:STUdef}
\end{eqnarray}
where $c\equiv m_W/m_Z$ and $s\equiv(1-c^2)^{1/2}$.
The choice of $\alpha$, $m_W$, and $m_Z$ as inputs is convenient for our purposes,
since their relation to the parameters in the Lagrangian is independent of how the fermions are incorporated into the theory.  Thus, $s$ and $c$, and therefore equations (\ref{eq:STUdef}) are independent of which model of fermions we are using.  Note that our choice of
definition for $\sin^2{\theta_W}$ is different from that used in Ref.~\cite{Burgess:1993vc}.

In the fermion model I, we obtain
\begin{eqnarray}
g^{CC(I)}_{L0}&=& {e\over s}\left[1+\lambda^2/6 +{\cal O}(\lambda^4)\right]\nonumber\\
g^{NC(I)}_{L0}&=& {e\over sc}\left[1+\lambda^2/6 +{\cal O}(\lambda^4)\right]\nonumber\\
g^{NC(I)}_{Q0}&=& -{es\over c}\left[1+{\cal O}(\lambda^4)\right]
\ .
\label{eq:couplings1}
\end{eqnarray}
Thus, we find in this theory
\begin{eqnarray}
\alpha S&=& 2s^2\lambda^2/3\nonumber\\
\alpha T&=&0\nonumber\\
\alpha U&=&0\ .
\label{eq:STU1}
\end{eqnarray}
This nonzero $S$ in fermion model I is problematic, because in order to make it small
enough to be compatible with data, the tower of vector boson states must be too heavy
to salvage the unitarity bound.

\paragraph{(ii) Couplings in Model II.}

Let us begin by assuming that $t_{\chi_R}=0$.
To the extent to which we can neglect $t_{\chi_R}$, the right-handed couplings
of the light fermions, $g^{CC}_{Rn}$ and $g^{NC}_{Rn}$, are also negligible. 
The remaining couplings of the light fermions in Model II can be expressed
\begin{eqnarray}
g^{CC(II)}_{Ln}&=&g^{CC(I)}_{Ln}  \int_0^{\pi R} dy\left[{1\over\pi R}+{1\over t_L^2}\delta(y)\right]
\left({f_n(y)\over f_n(0)}\right)\alpha_{u_0}(y)\alpha_{d_0}(y)\nonumber\\
g^{NC(II)}_{Ln}&=&g^{NC(I)}_{Ln}\int_0^{\pi R} dy\left[{1\over\pi R}+{1\over t_L^2}\delta(y)\right]
\left({g_n(y)-g_n(\pi R)\over g_n(0)-g_n(\pi R)}\right)\alpha_{\chi_0}(y)^2\nonumber\\
g^{NC(II)}_{Qn}&=& g^{NC(I)}_{Qn}\ ,
\label{eq:couplings2p}
\end{eqnarray}
For $t_{\chi_R}=0$, the left-handed
fermion wavefunctions within a doublet are identical, $\alpha_u=\alpha_d$.

If we assume that $t_L\approx\lambda$, the couplings of the light fermions to the heavier vector
bosons are suppressed by a factor of $m_W/m_{W_n}\approx \lambda/(n\pi)$, relative
to their couplings to the standard model $W$ and $Z$, just as in model I.  Although the bulk fermions have a larger overlap with the heavy vector bosons (of order $1/\lambda^2$),
this is compensated by the small probability for the light fermions to leak into the bulk (of order
$t_L^2$).   Thus, the only effect is that the expressions~(\ref{eq:CCKK1}) for model I
are multiplied by an overall coefficient of order $t_L^2/\lambda^2\sim1$.
As an example, for
$M=0$, the left-handed charged-current and neutral-current couplings are modified, for
$n$ odd, to
\begin{eqnarray}
g^{CC(II)}_{Ln}&=& g^{CC(I)}_{Ln}\left[1-{2t_L^2\over\lambda^2}+{\cal O}(\lambda^2,t^2)\right]\nonumber\\
g^{NC(II)}_{Ln}&=& g^{NC(I)}_{Ln}\left[1-{2t_L^2\over\lambda^2}
{g^2\over{g^2-g^{\prime2}}}
+{\cal O}(\lambda^2,t^2)\right]\ .
\label{eq:CCKKII}
\end{eqnarray}
The $g^{NC(II)}_{Ln}$ and $g^{CC(II)}_{Ln}$ couplings, for $n>0$ and even, are not affected by the delocalization at leading order, and the $g^{NC(II)}_{Qn}$ coupling is not affected
at any order.
Thus, the argument for ignoring the contributions of the heavy vector bosons to
four-fermion operators holds equally in model II, as it did in model I.

As for the couplings $g^{CC}_{L0}$ and $g^{NC}_{L0}$ of the light fermions to the standard model vector bosons, they are multiplied by a small suppression factor in model II, relative to the couplings in model I.
This is seen from the fact that
the fermion wave functions are normalized by
\begin{eqnarray}
1\ =\ \int_0^{\pi R} dy\left[{1\over\pi R}+{1\over t_L^2}\delta(y)\right]
\alpha_\chi (y)^2\ ,
\label{eq:fermnorm}
\end{eqnarray}
while the factors in parentheses in equations~(\ref{eq:couplings2p}) are positive and
less than one:
\begin{equation}
0\le\qquad {f_0(y)\over f_0(0)} \approx {g_0(y)-g_0(\pi R)\over g_0(0)-g_0(\pi R)}
\approx 1-{y\over\pi R}\qquad \le1\ .
\end{equation}
The suppression factors for $g^{CC}_{L0}$ and $g^{NC}_{L0}$ are identical to leading order in $\lambda^2$.

Evaluating the integrals, we obtain
\begin{eqnarray}
g^{CC(II)}_{L0}&=&g^{CC(I)}_{L0}  (1-At_L^2)\nonumber\\
g^{NC(II)}_{L0}&=&g^{NC(I)}_{L0}(1-At_L^2)\nonumber\\
g^{NC(II)}_{Q0}&=& g^{NC(I)}_{Q0}\ ,
\label{eq:couplings2}
\end{eqnarray}
where
\begin{eqnarray}
A&=&e^{-\hat{M}}{\sinh\hat{M}\over\hat{M}}
-{1\over2\hat{M}}\left(1-e^{-\hat{M}}{\sinh\hat{M}\over\hat{M}}\right)
\ .
\label{eq:CCsm}
\end{eqnarray}
In the limit $\hat{M}\rightarrow0$ we find $A\rightarrow1/2$ .
This was the main result of Ref.~\cite{Foadi:2004ps}.  By allowing the fermions to extend into the bulk, 
as in model II, one can cancel the effects of $S$ in electroweak measurements.  Comparing Eq.~(\ref{eq:couplings2})
with Eq.~(\ref{eq:couplings1}), we see that $S$ can effectively be set to zero (while retaining $T=U=0$) by the choice
\begin{equation}
t_L^2\ =\ {\lambda^2\over6A}\ .
\label{eq:leakage}
\end{equation}

Finally, let us now consider the effects of a small, but nonzero $t_{\chi_R}$.
A small $t_{\chi_R}$ will modify the couplings
$g^{CC}_{Ln}$, $g^{NC}_{Ln}$, and $g^{NC}_{Qn}$ only slightly, since 
it only contributes at order $t^4$ at the earliest.     The most important
effect will be the introduction of nonzero right-handed couplings for the light fermions:
\begin{eqnarray}
g^{CC(II)}_{Rn}&=& \int_0^{\pi R} dy{1\over\pi R}
f_n(y)\beta_{u_0}(y)\beta_{d_0}(y)\nonumber\\
g^{NC(II)}_{Rn}&=& \int_0^{\pi R} dy{1\over\pi R}
\Bigl(g_n(y)-g_n(\pi R)\Bigr)\beta_{\chi_0}(y)^2\ .
\label{eq:CCcouplingsR}
\end{eqnarray}
For small $t_R$, we note that $\beta_\chi (y)\sim{\cal O}(t_{\chi_R})$, so that the right-handed charged couplings will only be important for the third generation.  Up to corrections of order
$\lambda^2$ and $t^2$, we obtain for $M=0$ the following right-handed couplings between
the light fermions and the standard model $W$ and $Z$ bosons:
\begin{eqnarray}
g^{CC(II)}_{R0}&=& {g\over2}t_{u_R}t_{d_R}\left[1+{\cal O}(\lambda^2,t^2)\right]\nonumber\\
g^{NC(II)}_{R0}&=& {g\over2c}t_{\chi_R}^2\left[1+{\cal O}(\lambda^2,t^2)\right]\ ,
\label{eq:CCcouplingsReval}
\end{eqnarray}
where $c\equiv m_W/m_Z$.

\begin{figure}
\includegraphics[width=6in,height=4.8in]{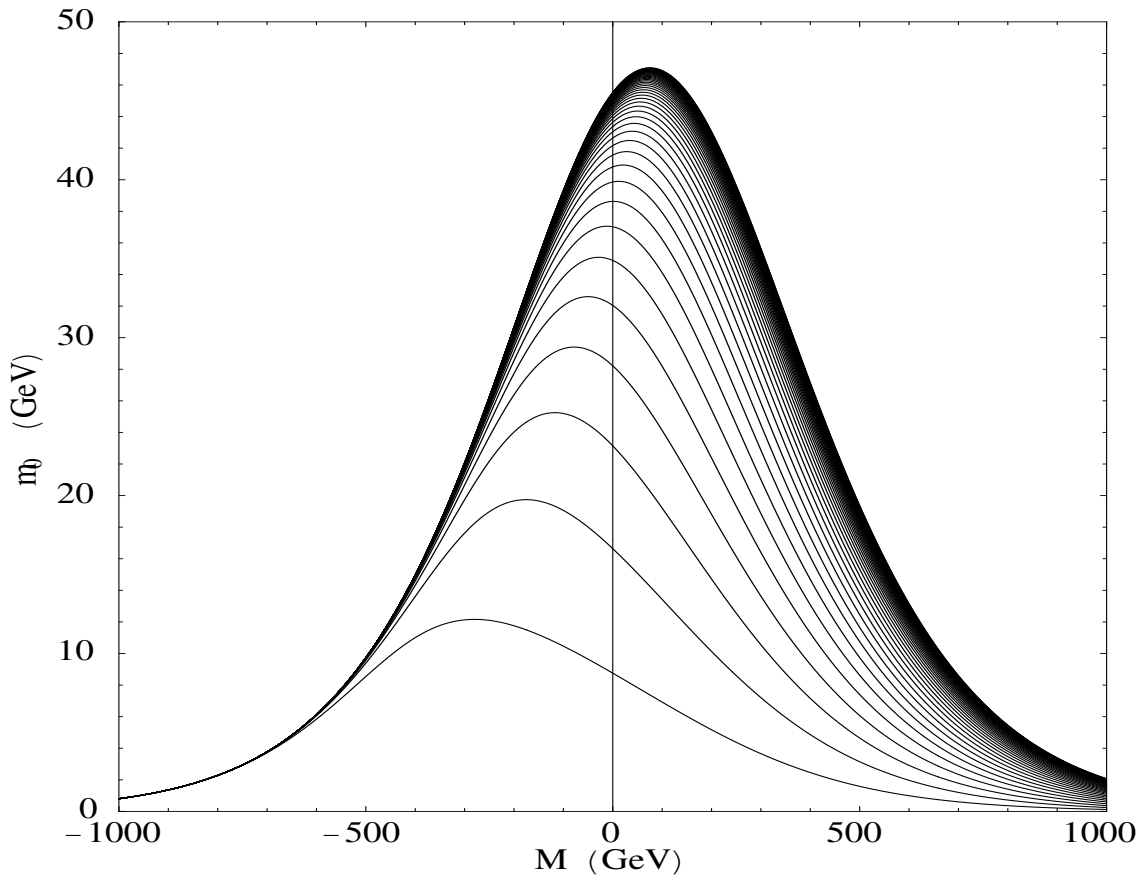}
\caption{Mass of the lightest fermion, as a function of the bulk mass, for $t_L$ chosen to adjust
the $S$-parameter to zero, and $1/R=500$ GeV. The curves correspond to several values of $t_{\chi_R}$,
from $10^{-1}$ to infinity.}
\label{fig:mass2}
\end{figure}

\section{The Top Quark}
\label{sec:top}

In  Section \ref{subsec:fermionmass} it was shown that the fermion masses are suppressed
by the factor $t_Lt_{\chi_R}$.  In Section \ref{sec:constraints}, we saw that we could 
choose $t_L$ to cancel gauge sector contributions to the
$S$ parameter, thereby relating $t_L$ to $\lambda$ by Eq.~(\ref{eq:leakage}).
Thus, we are left with $t_{\chi_R}$ as the final degree of freedom to fit the fermion masses.  
This works well for all of the light fermions, except the top quark.  

Recall that the overall mass scale is set by $1/R$, which is
required to be less than about a TeV in order to sufficiently delay unitarity in
$W_L^+W_L^-\rightarrow W_L^+W_L^-$ scattering.  Then with $t_L$ fixed to cancel the
contributions to the $S$ parameter, it is impossible to obtain a realistic top quark mass of
175 GeV.  For example, for $1/R=500$ GeV, $M=0$, and $t_L$ fixed by 
Eq.~(\ref{eq:leakage}), the lightest fermion mass solution to Eq.~(\ref{eq:massequation2})
has a maximum value of about 45 GeV.  Even if we allow the bulk mass $M$ to
be nonzero, we cannot do much better, since Eq.~(\ref{eq:leakage})
 involves $M$ in a dramatic way. In particular, when
$M\rightarrow -\infty$, $t_L$ tends exponentially to zero, and so does $m_0$.  In the
other limit of $M\rightarrow\infty$, the solution for $m_0$ itself is exponentially
suppressed, as shown in Eq.~(\ref{eq:mass2}).  Thus we find a 
peak near $M=0$ in the curve for $m_0$ as a function of $M$ (for fixed
$t_{\chi_R}$), as shown in Fig.~\ref{fig:mass2}.  From this curve with
$1/R=500$ GeV, we find that the maximum possible quark mass for any value of $M$
is about 47 GeV, which occurs for $t_{\chi_R}$ set to infinity.

One possible way to solve this problem is to allow a
different $t_L$ for the third generation of quarks. This approach might be viable, since the
constraints on the $S$ parameter do not directly involve the third generation fermions.
However, we find it unattractive, since universality of $t_L$ (and $M$)
was the simplest way to avoid any dangerous flavor-changing neutral currents.
An alternate solution that we prefer is to allow a different size of $R$ for the gauge sector
and the fermion sector.  This possibility had been suggested in the context of the
warped-space model in Ref.~\cite{Cacciapaglia:2004rb}.  It is even more sensible in
the context of a theory space, since there is no reason for the coefficients of
$\bar{\psi}\Gamma^{\mu} D_{\mu}\psi$ and $\bar{\psi}\Gamma^5 D_5\psi$ to be identical, in the bulk sector of
Eq.~(\ref{eq:fermionaction2}).  In terms of the 
deconstructed~\cite{Arkani-Hamed:2001ca, Hill:2000mu}
version of the theory,
this just corresponds to allowing the gauge couplings and the Yukawa couplings to be independent of each
other~\cite{Georgi:2005dm}.  The most general extension of model II, with
y-independent parameters, is described by the action ${\cal S}+{\cal S}^{(II)\prime}$,
where ${\cal S}$ is just the gauge sector action given in
Eq.~(\ref{eq:5daction}), and
\begin{eqnarray}
{\cal S}^{(II)\prime} = \int_0^{\pi R}dy && \int d^4x
\left[{1\over\pi R}\left(\bar{\psi}i\Gamma^{\mu} D_{\mu}\psi
+{\kappa}\left({1\over 2}\bar{\psi}i\Gamma^5 D_5\psi + \ {\rm h.c.}\right)
-M\bar{\psi}\psi\right)\right.\nonumber\\
&&\left.
+\delta(y){1\over t_L^2}i\bar{\psi}_L\sla{D}\psi_L
+\delta(\pi R-y)\left({1\over t_{u_R}^2}i\bar{u}_R\sla{D}u_R
+{1\over t_{d_R}^2}i\bar{d}_R\sla{D}d_R\right)
\right]\ ,
\label{eq:fermionaction3}
\end{eqnarray}
where $\kappa$ is a new parameter.\footnote{From an extra-dimensional point of view, this action
corresponds to a theory with a microscopic breaking of the five-dimensional Lorentz invariance,
in addition to the macroscopic breaking due to the compactification.} Rescaling the parameter $y$ by
$y\rightarrow y'=y/{\kappa}$, the action ${\cal S}^{(II)\prime}$ becomes identical to ${\cal S}^{(II)}$, with
$R$ replaced by $R/\kappa$.\footnote{The only difference is the interaction term with $W_5^a$, which is zero in unitary gauge anyway.}
Therefore, the gauge sector mass scale, $1/R_g\equiv 1/R$, and the
fermion sector mass scale, $1/R_f\equiv {\kappa}/R$, are independent quantities in theory space:
setting $R_f=R_g$ is an unnecessary and arbitrary choice.

With the action ${\cal S}^{(II)\prime}$ replacing ${\cal S}^{(II)}$, the fermion masses are (for $M=0$),
\begin{eqnarray}
\hat{m}_0 & = & \kappa {t_L t_{\chi_R}\over\sqrt{1+t_{\chi_R}^2}}\left[1-{1+t_{\chi_R}^2
+t_{\chi_R}^4/3\over 2\left(1+t_{\chi_R}^2\right)^2}t_L^2 + {\cal O}(t_L^4)\right] \label{eq:mass3} \\
\hat{m}_n & = & \kappa\pi\left(n-{1\over 2}\right)\left[1+{t_L^2+t_{\chi_R}^2\over\pi^2\left(n-{1\over 2}\right)^2}
+{\cal O}(t^4)\right];\ \ n=1,2,...\ . \label{eq:mass_n2}
\end{eqnarray}
Now we can account for $m_t$ by simply increasing $\kappa$.  Furthermore, it is now
possible to suppress the right-handed couplings in Eq.~(\ref{eq:CCcouplingsReval}), since
a larger $\kappa$ requires a smaller $t_{\chi_R}$ to obtain a given fermion mass.
Of course $\kappa$ (and $1/R_f$) cannot be too large, due to unitarity constraints
similar to those which
give bounds on $1/R_g$.  In the case of $1/R_f$ the limits come from
scattering processes such as $t\bar{t}\rightarrow W_L^+W_L^-$.  We shall investigate
these unitarity bounds in the next section.

\begin{figure}[!t]
\includegraphics[width=6in]{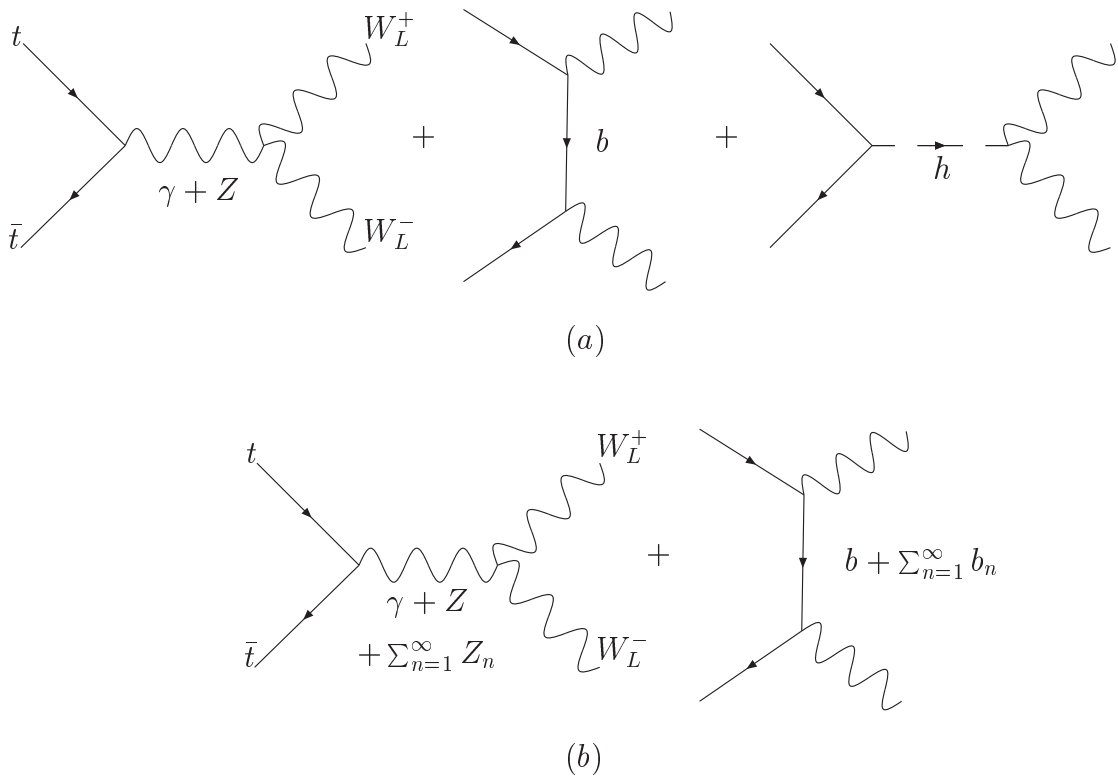}
\caption {(a) Diagrams contributing to the $t \bar{t} \rightarrow W^+_LW^-_L$ tree-level scattering amplitude in the SM. (b) Same process, in theory space.} \label{fig:amplitude}
\end{figure}

\section{Unitarity Constraints}
\label{sec:unitarity}

Here we shall restrict ourselves to considering the unitarity bounds coming from the
$t\bar{t}\rightarrow W_L^+W_L^-$ scattering process.  General constraints on couplings
in Higgsless models from this and related
processes have been considered previously in Ref.~\cite{Schwinn:2005qa}.  
In the SM, the tree-level $t\bar{t}\rightarrow W_L^+W_L^-$ scattering amplitude is given by the four
diagrams of Fig.~\ref{fig:amplitude}(a).  
If $t$ and
$\bar{t}$ have opposite helicities, the $\gamma$- and $Z$-exchange diagrams produce quadratically divergent terms, in the high-energy limit, which are 
cancelled by the $b$-exchange diagram~\cite{Lee:1977eg}.  The Higgs boson is not involved
in this cancellation, which is confined to the $J=1$ partial wave, so there is
no quadratic growth of the amplitude with energy, regardless of the Higgs boson mass.  
If $t$ and $\bar{t}$ have the 
same helicity, the $b$-exchange diagram produces a linearly
divergent high-energy term in the $J=0$ channel, which is cancelled by the Higgs boson exchange
diagram. 

In our Higgsless model the Higgs boson exchange diagram, of course, does not occur.
The $\gamma$-, $Z$-, and $b$-exchange diagrams are supplemented by corresponding
diagrams with exchange of heavy $Z_n$'s and $b_n$'s, as shown in Fig.~\ref{fig:amplitude}(b).
As $1/R_f\rightarrow\infty$, these heavy $Z_n$'s and $b_n$'s are removed from the theory,
and it becomes equivalent to the SM without the Higgs boson.  Thus, it is reasonable to
expect that the cancellation that occurs for opposite helicity $t$ and $\bar{t}$ in the SM
also occurs in
our Higgsless model, and that the amplitude does not display quadratic energy growth
at any scale.  We have directly verified this in our model.  However, if the $t$ and $\bar{t}$
have the same helicity, the linear growth in energy, that was cancelled by Higgs boson
exchange in the SM, now must be cancelled by some other sector of the theory.  
In our Higgsless model this cancellation occurs through the $b_n$-exchange diagrams.
In this respect, the heavy $b$-quarks play the role of the SM Higgs boson for this scattering
process.

In Ref.~\cite{Foadi:2003xa} we used the quadratic growth in energy of the
$W_L^+W_L^-\rightarrow W_L^+W_L^-$ scattering amplitude to place approximate bounds 
on the scale $1/R_g$, where the heavy vector states come in to restore unitarity.  We can 
now do the same here, using the $t \bar{t} \rightarrow W^+_LW^-_L$ process to place 
approximate bounds on $1/R_f$.  Note that, since the fermion amplitude
only shows linear growth with energy in the high-energy limit, the corresponding limits on the
heavy fermion states will be significantly weaker.

For left-handed $t$ and $\bar{t}$, the $J=0$ partial wave amplitude is given by
\begin{eqnarray}
& & a_0(t_L\bar{t}_L\rightarrow W_L^+W_L^-)=
{1\over 64\pi}\sum_{n=0}^{\infty}{1\over p^2+k^2+m_{b_n}^2}\nonumber \\
& & \left[
\left((h^{CC}_{Ln})^2+(h^{CC}_{Rn})^2\right)\left(m_t k g(\xi _n)
+2{m_t\over m_W^2}p E^2 h(\xi _n)-m_t p f(\xi _n)\right)\right. \nonumber\\
& & \left.+ {m_{b_n}\over m_W^2} h^{CC}_{Ln} h^{CC}_{Rn}\left(4 k E^2 g(\xi _n)
-2p(2E^2-m_W^2) f(\xi _n)\right)\right], \label{eq:amplitude}
\end{eqnarray}
where
\begin{equation}
\xi _n\equiv {2pk\over p^2+k^2+m_{b_n}^2}, \label{eq:xi}
\end{equation}
$E$ and $p$ are the $t$ (or $\bar{t}$) energy and momentum, respectively, 
$k$ is the $W_L^+$ (or $W_L^-$) momentum, 
and $m_{b_n}$ is the $b_n$ mass, with $m_{b_0}\equiv m_b$.
The functions $f(x)$, $g(x)$, and $h(x)$ are
\begin{eqnarray}
f(x) & = & {1\over x}{\ln {1+x\over 1-x}} \nonumber\\
g(x) & = & -{2\over x}\left(1-{1\over 2x}{\ln {1+x\over 1-x}} \right) \nonumber\\
h(x) & = & {2\over x^2}\left(1-{1-x^2\over 2x}{\ln {1+x\over 1-x}} \right). \label{eq:functions}
\end{eqnarray}
\begin{figure}
\includegraphics[width=6in,height=4.8in]{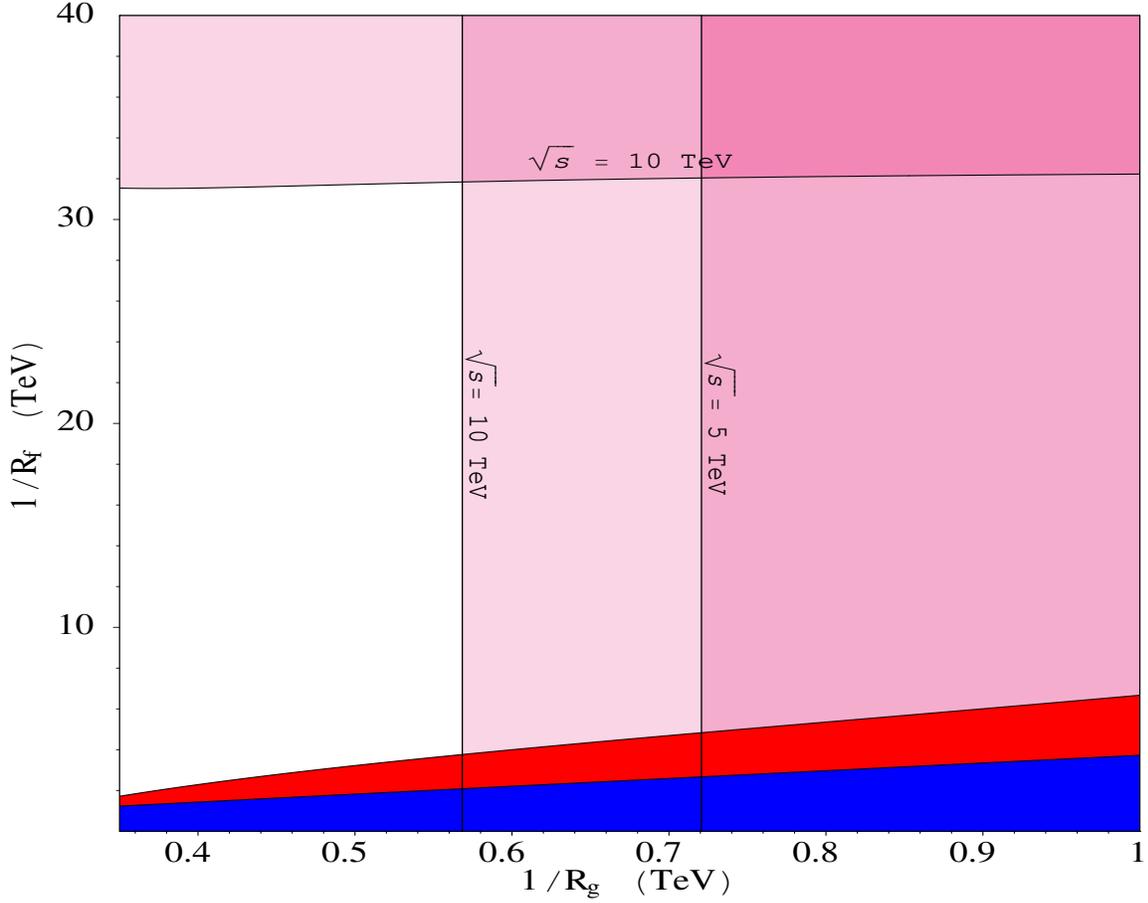}
\caption{Bounds imposed by unitarity constraints on the $t \bar{t}\rightarrow W^+_L W^-_L$ scattering at $\sqrt{s}=$10 TeV
(upper curve), and the $W^+_L W^-_L\rightarrow W^+_L W^-_L$ scattering at $\sqrt{s}=$10 TeV and $\sqrt{s}=$5 TeV (vertical lines),
in the ($1/R_g,1/R_f$) plane.
Specifically, we have assumed the requirement of
$a_0<1/2$ for both scattering processes. The $t \bar{t}\rightarrow W^+_L W^-_L$ scattering
at $\sqrt{s}=$5 TeV imposes no bound, since at this energy even the SM without Higgs boson does not violate unitarity.
The two curves on the bottom correspond to the minimum value of $1/R_f$ which allows a top mass
of 175 GeV to be a solution of the mass equation for $M=0$ (lower), and the minimum value of
$1/R_f$ which gives a $tbW$ right-handed coupling in agreement with the experimental constraint (upper).
} \label{fig:unitarity}
\end{figure}
The couplings $h^{CC}_{Ln}$ and $h^{CC}_{Rn}$ are obtained from 
the generalizations of Eqs.~(\ref{eq:couplings2p})
and (\ref{eq:CCcouplingsR}) by replacing $u\rightarrow t$, $d\rightarrow b_n$,
and $W_n\rightarrow W_0\equiv W$; that is\footnote{In this section we 
have omitted the superscript ``(II)'' on the gauge couplings.},
\begin{eqnarray}
h^{CC}_{Ln}&=& \int_0^{\pi R} dy\left[{1\over\pi R}+{1\over t_L^2}\delta(y)\right]
f_0(y)\alpha_{t_0}(y)\alpha_{b_n}(y)\nonumber\\
h^{CC}_{Rn}&=& \int_0^{\pi R} dy{1\over\pi R}
f_0(y)\beta_{t_0}(y)\beta_{b_n}(y)\nonumber\\
\label{eq:CCcouplingsGeneral}
\end{eqnarray}
For $M=0$, to leading order in $t_L^2$, $t_{b_R}^2$, and $\lambda^2$, they are
\begin{eqnarray}
h^{CC}_{L0} & = & g\left[1+{\cal O}(t^2)\right] \nonumber \\
h^{CC}_{Ln} & = & g{\sqrt{2}t_L\over 1+t_{t_R}^2}
\left[{(-1)^{n+1}\over \pi^2\left(n-{1\over 2}\right)^2}+{2t_{t_R}^2\over\pi^3\left(n-{1\over 2}\right)^3}\right]\left[1+{\cal O}(t^2)\right];\ \ n=1,2,... \nonumber \\
h^{CC}_{R0} & = & g{t_{t_R}t_{b_R}\over 2\sqrt{1+t_{t_R}^2}}\left[1+{\cal O}(t^2)\right] \nonumber \\
h^{CC}_{Rn} & = & g{\sqrt{2}\over \pi^2\left(n-{1\over 2}\right)^2}{t_{t_R}\over\sqrt{1+t_{t_R}^2}}
\left[1+{\cal O}(t^2)\right];\ \ n=1,2,...
\label{eq:coeff_M=0}
\end{eqnarray}

\begin{figure}
\includegraphics[width=6in,height=4.8in]{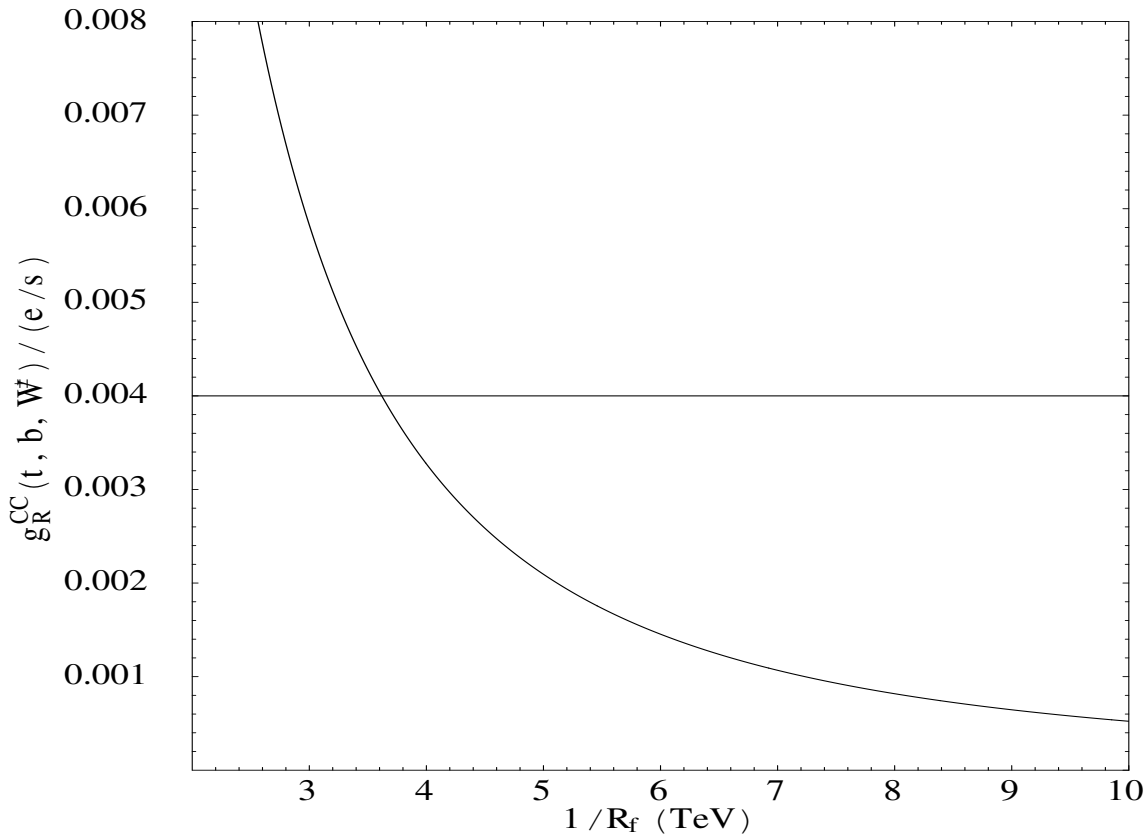}
\caption{Right-handed $tbW$ coupling, in units of $e/s$, for $1/R_g=550$ GeV, as a
function of $1/R_f$. The horizontal line corresponds to the experimental bound of Ref.~\cite{Larios:1999au}.
} \label{fig:RHCoupling}
\end{figure}

In the high energy limit, Eq.~(\ref{eq:amplitude}) becomes
\begin{equation}
a_0(t_L\bar{t}_L\rightarrow W_L^+W_L^-)\simeq {1\over 32\pi}{m_t E\over m_W^2}\sum_{n=0}^{\infty}\left(
(h^{CC}_{Ln})^2+(h^{CC}_{Rn})^2-2{m_{b_n}\over m_t}h^{CC}_{Ln}h^{CC}_{Rn}\right). \label{eq:HElimit}
\end{equation}
It is straightforward to show that this vanishes, to leading order in $t_L^2$, $t_{b_R}^2$, and $\lambda^2$, using the couplings given in Eq.~(\ref{eq:coeff_M=0}) and the masses given
in Eq.~(\ref{eq:mass3}) and Eq.~(\ref{eq:mass_n2}), applied to $m_t$ and $m_{b_n}$, respectively.
In fact, using the completeness relations
\begin{eqnarray}
\sum_{n=0}^{\infty}\left({1\over \pi R}+{\delta (y)\over t_L^2}\right)\alpha_{b_n}(y)\alpha_{b_n}(y')
& = & \delta (y-y') \nonumber \\
\sum_{n=0}^{\infty}\left({1\over \pi R}+{\delta (y)\over t_{b_R}^2}\right)\beta_{b_n}(y)\beta_{b_n}(y')
& = & \delta (y-y') \label{eq:completeness},
\end{eqnarray}
as well as the equations of motion, Eq.~(\ref{eq:wave}), and the boundary conditions, Eq.~(\ref{eq:BCwave}),
for the $t$, and the $b_n$'s, it can be shown that
\begin{equation}
\sum_{n=0}^{\infty}\left(
(h^{CC}_{Ln})^2+(h^{CC}_{Rn})^2-2{m_{b_n}\over m_t}h^{CC}_{Ln}h^{CC}_{Rn}\right)\equiv 0. \label{eq:vanishes}
\end{equation}
Therefore, this cancellation is exact in this model for any values of the couplings, and the 
linear growth in energy at high energies does not occur.

Of course, the cancellation of the term that grows with energy is not a sufficient condition for the unitarization of the
amplitude:  the latter could stop growing after unitarity is already violated. The heavy $b$-quarks should
come into play early enough to cancel the bad high-energy behavior, and this is only possible if $1/R_f$ is not too large.
Enforcing Eq.~(\ref{eq:leakage}), to keep $S$ fixed at zero, and setting $M=0$, the only parameters that are not fixed by
the light SM fermions and bosons are
$R_g$ and $R_f$, which set the scale for the heavy vector bosons and the heavy fermions, respectively.
We can put some reasonable constraints on these two parameters by requiring that
the $t_L\bar{t}_L\rightarrow W_L^+W_L^-$ and the $W_L^+W_L^-\rightarrow W_L^+W_L^-$
scattering amplitudes remain unitary up to some value of the center-of-mass energy
$\sqrt{s}$.  As an example, in Fig.~\ref{fig:unitarity} we display the region in the
($1/R_g,1/R_f$) plane that is allowed by the requirement that $a_0<1/2$ up to
$\sqrt{s}=$10 TeV or $\sqrt{s}=$5 TeV for both scattering 
amplitudes\footnote{The expression for the
$W_L^+W_L^-\rightarrow W_L^+W_L^-$ scattering amplitude can be obtained by taking
the $N\rightarrow\infty$ limit of Eq.~(3.7) in Ref.~\cite{Foadi:2003xa}.   }. 
As expected, we see that $1/R_f$ can indeed be much larger
than $1/R_g$. If we require the theory to respect unitarity up to $\sqrt{s}=$10 TeV in both amplitudes,
we find $1/R_g\lesssim570$ GeV and
$1/R_f\lesssim32$ TeV.  If we use the weaker requirement that the theory only respect
unitarity up to $\sqrt{s}=$5 TeV, then we find $1/R_g\lesssim720$ GeV, while there
is no constraint on $1/R_f$,  since the
$t_L\bar{t}_L\rightarrow W_L^+W_L^-$ scattering amplitude
does not violate unitarity at this energy even in the SM without a Higgs boson.
Of course, any bounds on $1/R_g$ and $1/R_f$ depend on the somewhat arbitrary 
scale choice for $\sqrt{s}$, where the low energy Higgsless theory has broken down.

Lower limits on the scales $1/R_g$ and $1/R_f$ can be obtained from experimental input.  In
Ref.~\cite{Foadi:2003xa} it was estimated that $1/R_g\gtrsim500$ GeV
from limits on $W^\prime$ and $Z^\prime$ states with appropriate couplings.
This lower limit should be even weaker in model II, with delocalized fermions, since
the coupling between the light fermions and the heavy vector states have an additional, order one,
suppression relative to that in model I, as given in Eqs.~(\ref{eq:CCKKII}).
For the case of $1/R_f$, a minimal requirement is that it be large enough to accommodate
a top quark mass of 175 GeV.  This is displayed in the lower curve on the bottom of Fig.~\ref{fig:unitarity}.
It gives a lower bound of $1/R_f\gtrsim1-3$ TeV, with the dependence on $1/R_g$ entering
through the condition imposed by Eq.~(\ref{eq:leakage}).  However,
this curve corresponds to an infinite value of $t_{t_R}$, which is not viable.  Tighter constraints
can be obtained by limits on the right-handed $tbW$ and $ttZ$ couplings, which are 
evaluated to lowest order in $t_{\chi_R}$ in
Eqs.~(\ref{eq:CCcouplingsReval}).
For example, in Ref.~\cite{Larios:1999au} it is estimated, using experimental
results on the $b\rightarrow s\gamma$ process, that
$g^{CC}_{R}(t,b,W^{\pm})/g\leq 0.4\cdot 10^{-2}$, at the $2\sigma$ level.  The corresponding 
bound on $1/R_f$ is displayed in the upper curve on the bottom of Fig.~\ref{fig:unitarity}.
For the particular value of $1/R_g=550$ GeV, we can see how the coupling
$g^{CC}_{R}(t,b,W^{\pm})/g$ (where we have used $g\equiv e/s$) varies with $1/R_f$ in Fig.~\ref{fig:RHCoupling}.  The experimental bound is satisfied for this value of $1/R_g$ by $1/R_f\gtrsim3.6$~TeV, which corresponds to $\kappa\gtrsim6.5$. 
An even stronger bound might be obtainable from limits
on the right-handed neutral current coupling, since it is quadratic in the parameter $t_{t_R}$;
however, the extraction of this coupling requires more detailed analysis of higher order
effects at the $Z$-pole in our model.

\section{Conclusions}
\label{sec:conclusions}

In this paper we have completed the construction of a phenomenologically viable higgsless model from theory space,
by extending and generalizing the models of Refs.~\cite{Foadi:2003xa} and~\cite{Foadi:2004ps}.
This model contains three features, which are crucial to any viable higgsless model of electroweak symmetry breaking.  First, it contains a tower of vector bosons which delay the unitarity violation in the $W_L W_L\rightarrow W_L W_L$ and $W_L Z_L\rightarrow W_L Z_L$ scattering
amplitudes, while giving the correct mass for the standard model $W^\pm$ and $Z$ (and photon)
as the lightest states in the tower.  Thus, it can extend the applicability of the effective higgsless
theory up to a higher scale in the 5-10 TeV range.  This is accomplished using an $SU(2)$ 
gauge symmetry on a theory-space interval, broken down to $U(1)$ at the right end of the interval, and with gauge kinetic terms on each end of the interval.  
The normalization ($\lambda$, $\lambda^\prime$) of the gauge kinetic terms on the boundaries
are easily arranged to give the correct mass for the SM $W^\pm$ and $Z$ bosons.

Second,  it incorporates a cancellation of the large vector boson contributions to the $S$ parameter, which generically occur in higgsless models.   This cancellation is obtained by allowing the light 
fermion wave functions to leak away from the ends of the interval.  In our model this leakage arises through boundary conditions and boundary kinetic terms for the fermions, where the light left-handed fields are predominantly located
at the left end of the interval and the right-handed fields are predominantly located at the right
end of the interval.  The leakage of the left-handed fields into the bulk can be made to cancel
the gauge boson contributions to $S$, while keeping the $T$ and $U$ parameters
naturally suppressed, by tuning the normalization ($t_L$) of the universal left-handed fermion kinetic term on the left boundary.  Meanwhile the normalization ($t_{\chi_R}$) of the right-handed fermion kinetic terms on the right boundary can be used to give the correct mass for each of the light fermions.
Furthermore, multiple generations and fermion mixings are implemented in the model, without introducing flavor-changing neutral currents, by confining all flavor physics to the right-handed fermion brane kinetic terms, and imposing a global $U(3)_{quark}\times U(3)_{lepton}$ symmetry on the bulk and left brane.

Third, it has a realistic top quark mass and small
nonstandard right-handed top and bottom couplings.  To obtain this goal, while maintaining the good unitarity properties of the $W_L W_L$ scattering,
it was necessary to separate the overall gauge sector scale ($1/R_g$) from the overall fermion sector scale ($1/R_f$).  This requires an explicit breaking of the five-dimensional Lorentz symmetry, which is theoretically allowed, since such symmetry is already
broken by compactification and brane kinetic terms.  In fact, within a theory-space model it can be considered natural, since the difference in 
the size of the scales is analogous to having different sizes of gauge and Yukawa couplings.
By making $1/R_f$ larger than $1/R_g$, it is possible to obtain the top quark mass.  It is also possible to suppress any nonstandard right-handed top and bottom couplings, since for a fixed
fermion mass, an increase in $1/R_f$ requires a compensatory decrease in $t_{\chi_R}$, leading
to a decrease in right-handed couplings.  

In this way, we have constructed a viable higgsless model with only three undetermined parameters, $1/R_g$, $1/R_f$, and the bulk fermion mass $M$.  Since the bulk fermion mass does not seem to add any qualitatively new features to the model, it is reasonable to set $M=0$, leaving
us with a two-parameter model.  
The parameter $1/R_g$ sets the scale of the vector boson excitations, and
the parameter $1/R_f$ sets the scale of the fermionic excitations.  Just as the scale $1/R_g$
cannot be too large and still effectively delay unitarity violation in $W_L W_L\rightarrow W_L W_L$
scattering, the scale $1/R_f$ cannot be too large and still effectively delay unitarity violation
in $t \bar{t}\rightarrow W_L W_L$ scattering.  Thus, both of these scales are bounded from above, the exact bounds depending on the energy scale at which the effective higgsless theory must be replaced by a more complete theory.  A reasonable upper bound for $1/R_g$ is in the 570-720 GeV range,
while the upper bound for $1/R_f$ is much weaker, of order 30 TeV or more.  Experimental lower limits on $1/R_f$ from right-handed $tbW$ couplings are in the range of 2-4 TeV.  Precise experimental lower limits on $1/R_g$ require further investigation, although given the small couplings between the light fermions and the heavy $W^\prime $ and $Z^\prime$ states, there appears to be a reasonable range for this parameter that is still allowed.

In conclusion, we have presented an existence proof of a viable higgsless model, that can
satisfy all current experimental constraints, as far as we know.  It is certainly not the only
higgsless model that may work, and it is probably too simplistic in many regards, but it has
all of the features that any higgsless model must have.  Thus, it offers a concrete example for use to explore the phenomenology of higgsless models at the Tevatron and the LHC.  In particular, it is worthwhile to further investigate its most relevant phenomenological aspects, with careful attention to those features which are general, rather than characteristic of any particular model.

\vskip .2 cm

\section*{Acknowledgments}

This work was supported by the US National
Science Foundation under grant PHY-0244789.


\end{document}